\documentclass[twocolumn,pre,superscriptaddress,a4paper,sort]{revtex4}
\usepackage{epsfig}
\newcommand{\kB}{k_{\mathrm{B}}}
\newcommand{\kT}{\kB T}
\newcommand{\lB}{l_{\mathrm{B}}}
\newcommand{\dens}{\rho}

\newcommand{\rhop}{\dens_+}
\newcommand{\rhom}{\dens_-}
\newcommand{\rhosalt}{\dens_s}
\newcommand{\rhoprot}{\dens_p}
\newcommand{\rhopm}{\dens_\pm}

\newcommand{\chempot}{\mu}

\newcommand{\mupm}{\chempot_\pm}
\newcommand{\kap}{\kappa}
\newcommand{\kappas}{\kap_s}
\newcommand{\phid}{\phi_{\mathrm{D}}}
\newcommand{\surftens}{\gamma}
\newcommand{\dsurftens}{\Delta\surftens_{\mathrm{el}}}
\newcommand{\dsurfsol}{\Delta\surftens_{\mathrm{s}}}
\newcommand{\dsurfxtal}{\Delta\surftens_{\mathrm{x}}}

\newcommand{\Molar}{\mathrm{M}}

\newcommand{\PB}{Poisson-Boltzmann}

\begin{document}

\title{Protein crystals and charged surfaces: interactions and
heterogeneous nucleation
}

\author{R. P. Sear}

\email{r.sear@surrey.ac.uk}

\affiliation{Department of Physics, University of Surrey, Guildford,
Surrey GU2 7XH, United Kingdom}

\begin{abstract}
As proteins typically have charges of around 10,
they will interact strongly with charged surfaces. We calculate the
electrostatic contribution to the interaction of crystals of protein
with charged surfaces. The surfaces repel like-charged crystals
and attract oppositely-charged crystals, with free energies which can
be easily several kT per protein molecule brought into contact with
the surface. This means that oppositely charged surfaces can act
as a nucleant, they can induce nucleation of a protein crystal by
lowering the free energy barrier to heterogeneous nucleation
of the crystal from a dilute solution.
\end{abstract}

\maketitle

\section{Introduction}

Here we will consider the interaction between charged surfaces and
protein crystals. Proteins are themselves charged and so we would
expect them to interact strongly with a surface that is charged.
Using the \PB~equation we find that indeed protein crystals are
attracted by oppositely charged surfaces and repelled by surfaces
with charges of the same sign. In itself this is hardly surprising
but with quite simple calculations we quantify this attraction and
repulsion. We show that easily accessible surface charge densities
of the opposite sign to that of the protein molecules,
are able to greatly favour heterogeneous nucleation at the
surface by reducing the free energy of a nucleus at the surface by
several $\kT$ per protein molecule at the surface. Controlled
heterogeneous nucleation is vital to the production of protein crystals,
which are required for structure determination via X-ray
crystallography.

At a first-order phase transition, such as the crystallisation of a protein
from solution, the transition
starts with nucleation \cite{debenedetti,rosenberger96}.
Protein crystallisation is the ``main bottleneck''
\cite{chayen02} in the
determination of the three-dimensional structure of proteins. Determining
this structure is crucial for understanding
what a protein does and how it does it.
Protein crystallographers wish to: i) {\em induce} nucleation in the relatively
weakly supersaturated solutions, within which protein crystals grow
slowly and so incorporate few defects, and ii) {\em control} heterogeneous
nucleation, in particular have nucleation occurring only at specific
locations on a surface. If crystals only nucleate on widely separated
patches on a surface then the growth of one crystal will not interfere with
and limit the growth of another crystal.

See for example Refs.~\cite{mcpherson88,chayen01} for experimental
work on adding solid surfaces to protein solutions to induce nucleation,
and see Refs.~\cite{sanjoh99,sanjoh01} for experimental work on
patterned surfaces, which shows that protein crystals can be made
to grow preferentially on specified patches of a surface. In
Refs.~\cite{sanjoh99,sanjoh01} a patterned surface is used in which
patches of the surface are one form of doped Si and the rest is
a different form. The protein nucleates preferentially on one of these
forms and so by controlling where this form is found in the pattern, the
crystals can be made to nucleate and grow far from each other, facilitating
the formation of large crystals. Here we show that a pattern of two
different surface types of opposite charges can produce sufficiently
large differences in surface free energy to make heterogeneous
nucleation much easier on one surface than on the other.

We note that here we only consider the effects of charge interactions,
screened by a 1:1 electrolyte.
%It is highly unlikely that all the experimental observations of
%Refs.~\cite{mcpherson88,sanjoh99,sanjoh01} are a result
%of simple electrostatics.
Other interactions, such as short-range specific interactions between
the protein molecules and surface
are known to be important, see the review
of Ostuni {\em et al.} \cite{ostuni99}. When these interactions
are important, our calculations, which neglect them, will only
yield estimates of the {\em variation}
with 1:1 salt concentration, of the interaction free energy
of a protein crystal or crystalline nucleus with a surface. However,
experimental systems do exist with high, variable
charges and highly hydrophilic surfaces. These highly hydrophilic
surfaces will have only weak non-electrostatic interactions with the proteins,
they also has the advantage of minimising the problem of protein denaturation.

One example is a highly charged
self-assembled monolayer, such as has been shown
to adsorb polyelectrolytes \cite{clark97}, another
is a membrane containing charged
lipids \cite{wagner00,murray99}.
Both of these systems should be well
described by the theory presented here.
There is a great deal
of work on the association of proteins and DNA with oppositely charged
lipid membranes, Refs.~\cite{wagner00,murray99} are just two examples
of a large literature.
As far as the author is aware, there has been no experimental study
of adsorption of protein on a highly charged, highly
hydrophilic, self-assembled monolayer, but such a system would offer
the opportunity to study the effect of electrostatics with minimal interference
from other interactions. This is true both for the adsorption of single
proteins and of protein crystals.

\begin{figure}
\begin{center}
\caption{
\lineskip 2pt
\lineskiplimit 2pt
A schematic of a nucleus of a protein crystal against a charged
surface. The hatched circles are the proteins in the crystal lattice,
each with its charge $Q$ marked. The surface is shaded grey and the
surface charges, positive in this case, are shown as $+$'s.
\label{surfxtal}
}
\vspace*{0.1in}
\epsfig{file=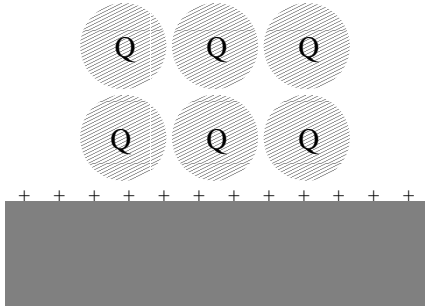,width=2.2in}
\end{center}
\end{figure}

Our calculations will all be for infinite planar interfaces, despite
the fact that our primary interest is in the nuclei of protein crystals
against charged surfaces, not bulk crystals against surfaces. Although,
nuclei are small, 10s of protein molecules, the part of the free energy
which scales with the contact area between the nucleus and the surface will
still dominate the edge contribution, as the Debye length is typically a few
nms at most. Thus, we consider only the dominant surface term,
considering sub-dominant terms within simple theories of simple generic models
is not useful. We will also leave consideration of defects in the
surface to later work.

In the next section we calculate the electrostatic contributions
to the surface free energy.
The third section includes example results and a discussion and the last
section is a conclusion.

\section{Calculation}

Here we calculate the electrical double layer contributions to the
interfacial free energies of solid--salt-solution, solid--protein-crystal
and solution--protein-crystal interfaces. The calculation for the
solution--protein-crystal interface is just a repeat of that done in
Ref.~\cite{sear02} and so will not be described in detail. We will use
the model of Ref.~\cite{sear02} for the protein crystal and solution,
along with a simple model of a charged solid surface. As we are interested in
interfacial free energies it is most convenient to work at constant
chemical potential of the salt, and so the appropriate free energy
is the grand potential $\Omega$. We will split this grand potential
into the part due to the charges on the surface and on the protein
molecules, together with the salt and counter ions, $\Omega_{el}$, and another
part which includes the rest of the interactions, $\Omega_{nel}$,
\begin{equation}
\Omega=\Omega_{nel}+\Omega_{el}.
\end{equation}
$\Omega_{nel}$ includes effects such as the dispersion interactions between
the protein molecules and the surface, and any short-range
interactions between the surfaces of the proteins and the solid surface.
%See Ref.~\cite{warren02} for a discussion of
%models of proteins with both short-range interactions and interactions
%due to the charge on the protein.
We will not calculate $\Omega_{nel}$,
and so will only be able to estimate the absolute values of the interfacial
free energies, when $\Omega_{nel}$ is much smaller than $\Omega_{el}$,
for example with highly charged, highly hydrophilic, surfaces, and
highly charged proteins.
However, if we assume that $\Omega_{nel}$ is weakly dependent
on the salt concentration, and that the charges on the surface can
be varied without varying $\Omega_{nel}$ significantly, our results
will describe the variation in the interfacial free energies with
surface charge density and the salt concentration.

Rather than considering a specific protein crystal, we
take over the jellium concept from the theory of metals to make a
general estimate of the effect of salt on the interfacial tension.  We
replace the detailed charge density due to the protein molecules by a
uniform background charge density $\rhoprot$, cut off abruptly at the
interface. In Ref.~\cite{sear02}, Sear and Warren did so for the
protein-crystal--salt-solution interface, and this followed
on from the work of Warren \cite{warren02} on the bulk phase behaviour.
Here when we
calculate the interfacial free energy for a protein crystal
at the surface, we will assume the
protein crystal is flush against the surface and so the
charge density is uniform right up to the surface.
%Our model of the charged surface of the substrate is that of
%an infinite charged plane at a fixed potential $\phi_s$.
Our model of the charged surface of the substrate is that of
an infinite charged plane at a fixed charge per unit area $\sigma$.
The solution outside the crystal is taken to be sufficiently
dilute that we can neglect any protein molecules in it and treat
it as a salt solution at a concentration $\rhosalt$.
For definiteness we take the protein to be positively charged so
the counterions are negative and the coions are positive; both are
monovalent. The surface is in the $xy$ plane at $z=0$, with
the solid in the $z<0$ half-space and either
a salt solution or a protein crystal in the $z>0$ half-space.

We will use the same notation as in Ref.~\cite{sear02}, so
we work in units where $e=\kT=1$.  In
these units, the Coulomb potential energy $U$ between a pair of
elementary charges separated by $r$ is $U=\lB/r$, where $\lB$ is the
Bjerrum length, equal to $0.72\,\mathrm{nm}$ in water at room
temperature ($\lB=e^2/4\pi\epsilon\kT$).  We assume a constant value
of $\lB$, and ignore dielectric effects. The
counterion and coion densities are $\rhom(z)$ and
$\rhop(z)$, respectively. In the bulk salt solution, both tend
to the salt concentration $\rhosalt$, and inside the
protein crystal $(\rhom-\rhop)\to\rhoprot$.

We will use a grand potential, $\Omega_{el}$, which contains only
ideal solution terms for the ions and the associated electric field at
the interface.
%We use a grand potential because as usual the
%calculation of the interfacial tension is easiest at fixed chemical
%potential not density.
The grand potential $\Omega_{el}$, is \cite{sear02},
\begin{equation}
\Omega_{el}=\int_{0}^\infty\!dz\,\omega(z),\quad
\omega=\sum_{i=\pm}\dens_i(\ln\frac{\dens_i}{\rhosalt}-1)
+\frac{E^2}{8\pi\lB}.\label{oeq}
\end{equation}
The first terms in $\omega$ are the ideal solution terms (the ions
share a common chemical potential $\mupm=\ln\rhosalt$).  The last
term is the electrostatic energy, wherein $E=-d\phi/dz$ is the
electric field strength corresponding to an electrostatic potential
$\phi$ which satisfies the Poisson equation,
\begin{equation}
\frac{d^2\phi}{dz^2}=
\left\{\begin{array}{ll}
-4\pi\lB\left(\rhop-\rhom\right)&~~~\mbox{salt solution}\\
-4\pi\lB\left(\rhop-\rhom+\rhoprot\right)&~~~\mbox{protein crystal}
\end{array}\right. ,
\end{equation}
The variational principle $\delta\Omega_{el}/\delta[\rhopm(z)]=0$
applied in this problem yields $\rhopm(z)=\rhosalt\exp[\mp\phi]$.  The
electrostatic potential then satisfies the \PB\ equation,
\begin{equation}
\frac{d^2\phi}{dz^2}-\kappas^2\sinh\phi = 
\left\{\begin{array}{ll}
0&~~~\mbox{salt solution}\\
-4\pi\lB\rhoprot&~~~\mbox{protein crystal}
\end{array}\right. , \label{pbeq}
\end{equation}
where 
\begin{equation}
\kappas^2=8\pi\lB\rhosalt.\label{kseq}
\end{equation}
For our assumption of a planar surface in the $z=0$ $xy$ plane, with
a fixed charge per unit area $\sigma$, the boundary condition on $\phi$
at $z=0$, is that the electric field at the surface
\begin{equation}
E=-\frac{ d\phi}{ dz}=\frac{\sigma}{\epsilon}=4\pi\lB\sigma.
\label{z0bc}
\end{equation}
The other boundary condition is that $d\phi/dz=0$,
$z\to\infty$.

\subsection{Surface--salt-solution interface}

This is just the problem addressed by Gouy and Chapman at the beginning
of the last century \cite{israelachvili}. For simplicity we will
only derive the expressions in the regime where the equations may
be linearised. Then we will simply state the general result obtained
when the equations are not linearised. Linearising the Poisson-Boltzmann
equation, Eq.~(\ref{pbeq}), inside the salt solution, and using
the boundary conditions, we have
\begin{equation}
\phi(z)=4\pi\lB\sigma\kappas^{-1}\exp(-\kappas z)~~~z>0.
\label{surfsalt}
\end{equation}
A simple exponential decay from a value at the surface, proportional
to the surface charge.

The surface free energy is, by definition, the difference between the actual
grand potential per unit area of the
surface--salt-solution interface and that it would
have if the salt solution continued unperturbed
right up to solid surface.
We therefore have to calculate the grand potential, Eq.~(\ref{oeq}), then
subtract the grand potential for the bulk solution:
Eq.~(\ref{oeq}) with the actual $\omega(z)$ replaced by its
value in the bulk salt solution, $\omega(+\infty)$ \cite{note},
\begin{equation}
\dsurfsol=\int_{0}^\infty\!dz\,\left(\omega(z)-\omega(+\infty)\right).
\label{oeq2}
\end{equation}

This is a general expression, in the linear regime we can expand out
$\omega$ of Eq.~(\ref{oeq}) and keeping only the terms up to
quadratic order in $\phi$, we have that $\omega=-2\rhosalt+2\rhosalt\phi^2$.
The bulk value  $\omega(+\infty)=-2\rhosalt$, as the potential in the
bulk of the salt solution is taken to be zero.
The surface free energy is then obtained by substituting these expressions
into Eq.~(\ref{oeq2}). Doing this and
using Eq.~(\ref{surfsalt}), we have
\begin{equation}
\dsurfsol=2\pi\lB\sigma^2\kappas^{-1}~~~\mbox{linear},
\label{gams}
\end{equation}
which is positive and proportional to the square of the surface
charge and to the screening length, as we might have expected.

If Eq.~(\ref{pbeq}) is not linearised the potential $\phi$ is given
by \cite{israelachvili,sear02} $\phi = 2\ln[(1+Ce^{-\kappas
z})/(1-Ce^{-\kappas z})]$ where $C$ is determined by the boundary
condition Eq.~(\ref{z0bc}) and is given by
\begin{equation}
C=\frac{\kappas}{2\pi\lB\sigma}\left[
\sqrt{1+\left(\frac{2\pi\lB\sigma}{\kappas}\right)^2}-1\right].
\label{csol}
\end{equation}
Thus, we can evaluate the surface free energy once we have converted
Eq.~(\ref{oeq2}) into an expression in terms of $\phi$ only.
Equation (\ref{pbeq}) in the salt solution can be integrated once
with respect to $\phi$,
\begin{equation}
\left(\frac{d\phi}{dz}\right)^2=
2\kappas^2\left(\cosh\phi-1\right),
\label{pbeqint1}
\end{equation}
which gives us the electric field term in $\omega$. The other, ideal gas, term
can be written in terms of $\rhosalt$ and $\phi$ using
$\rhopm(z)=\rhosalt\exp[\mp\phi]$.
Then $\dsurfsol$ is given by the integral
\begin{equation}
\dsurfsol=2\rhosalt\int_0^{\infty}\!dz \phi\sinh\phi,
\label{gamsol}
\end{equation}
which is straightforward to evaluate numerically.

\subsection{Surface--protein-crystal interface}

As with the surface--salt-solution interface we start with the
linear Poisson-Boltzmann equation. Here the potential is that
inside the protein crystal so we linearise Eq.~(\ref{pbeq}) and obtain
\begin{equation}
\frac{d^2\phi}{dz^2}=\kappas^2\left(\phi-\frac{\rhoprot}{2\rhosalt}\right),
\label{pblp}
\end{equation}
where we used Eq.~(\ref{kseq}) for $\kappas$. The second
term inside the parentheses is minus the Donnan potential $\phid$ inside the
protein crystal. The general expression for the Donnan potential is
\begin{equation}
\sinh\phid=\frac{\rhoprot}{2\rhosalt}\label{doneq},
\end{equation}
which when linearised gives $\phid=\rhoprot/(2\rhosalt)$. Thus,
the solution to Eq.~(\ref{pblp}), with the boundary
condition of fixed charge density,
Eq.~(\ref{z0bc}), is
\begin{equation}
\phi(z)=\phid+4\pi\lB\sigma\kappas^{-1}\exp(-\kappas z)~~~z>0.
\label{surfprot}
\end{equation}

In the bulk protein crystal, $\phi=\phid$. We expand
the $\omega$ of Eq.~(\ref{oeq}) in powers of $\phid$ and keep only the terms up to
quadratic order, and find that in the crystal
$\omega(-\infty)=-2\rhosalt +2\rhosalt\phid^2$ \cite{sear02}.
Performing the same expansion at the interface, we obtain
for the excess grand potential at the interface
\begin{equation}
\omega-\omega(-\infty)=
\rhosalt\left[\phi^2-\phid^2+2(\phi-\phid)^2\right],
\label{exox}
\end{equation}
where we used Eq.~(\ref{surfprot}) to obtain the derivative
$d\phi/dz$.
The interfacial free energy per unit area $\dsurfxtal$ is the
integral over all $z$ of the excess grand potential of Eq.~(\ref{exox}).
It is
\begin{equation}
\dsurfxtal=2\pi\lB\sigma^2\kappas^{-1}+\sigma\phid~~~\mbox{linear},
\label{gamp}
\end{equation}

If we do not linearise the Poisson-Boltzmann equation, Eq.~(\ref{pbeq}),
we can still integrate it once to obtain
\begin{equation}
\left(\frac{d\phi}{dz}\right)^2=
2\kappas^2\left(\cosh\phi-\cosh\phid\right)
-8\pi\lB\rhoprot\left(\phi-\phid\right).
\label{pbeqint}
\end{equation}
Combining this with the boundary condition, Eq.~(\ref{z0bc})
we obtain an equation for the potential at the surface $\phi(z=0)$,
\begin{eqnarray}
&&
\left[2\kappas^2\left(\cosh\phi(z=0)-\cosh\phid\right)
-8\pi\lB\rhoprot\left(\phi(z=0)-\phid\right)\right]^{1/2}\nonumber\\
&&=\pm4\pi\lB\sigma,
\label{z0pot}
\end{eqnarray}
which can be solved for $\phi(z=0)$, and then once this is known
the profile $\phi(z)$ is readily obtained by numerically
integrating Eq.~(\ref{pbeqint}). On the right hand side the $+$ ($-$) sign
is taken when $\sigma>0$ ($\sigma<0$).
%as then the gradient $d\phi/dz$
%(the left hand side of Eq.~(\ref{z0pot}))
%is the negative (positive) root of
%Eq.~(\ref{pbeqint}).
The interfacial free energy is then obtained
from
\begin{equation}
\dsurfxtal=\int_{0}^\infty\!dz\,\left(\omega(z)-\omega(-\infty)\right),
\label{oeq3}
\end{equation}
where the grand potential per unit volume in the protein crystal
is $\omega(-\infty)=-2\rhosalt(\cosh\phid-\phid\sinh\phid)$.
Using $\rhopm(z)=\rhosalt\exp[\mp\phi]$, and
Eq.~(\ref{pbeqint}) for the electric field term in Eq.~(\ref{oeq}), we have that
the electrical double layer
contribution to the free energy of the solid-protein interface is given by
\begin{equation}
\dsurfxtal=2\rhosalt\int_{0}^\infty\!dz\,
\phi[\sinh\phi-\sinh\phid].\label{dseq}
\end{equation}

\subsection{Salt-solution--protein-crystal interface}

This interface was the subject of Ref.~\cite{sear02}. In the
high salt regime where the Poisson-Boltzmann equation can be linearised
the electrical double layer contribution to the free energy
of the interface between a protein crystal and a salt solution is
\begin{equation}
\dsurftens=-\frac{\rhosalt\phid^2}{2\kappas}~~~\mbox{linear}.
\label{xtalsol}
\end{equation}
See Ref.~\cite{sear02} for the full nonlinear calculation.

\begin{figure}
\begin{center}
\caption{
\lineskip 2pt
\lineskiplimit 2pt
The electrical double layer contribution to the change
in free energy per unit area, times the surface area of the crystal per protein
molecule, when a protein
crystal is brought from the bulk of the salt
solution to the charged surface of a solid, $\Delta l^2$, Eq.~(\ref{delta}).
The solid, dotted, dot-dashed and dashed
curves are for charge densities $\sigma{e}=1$, $0.1$,
$-0.1$ and $-1~\!$nm$^{-2}$, respectively. Results are shown
for salt concentrations down to $0.025\Molar$. In the limit
of the salt concentration tending to zero the surface free energies
diverge.
\label{d}
}
\vspace*{0.2in}
\epsfig{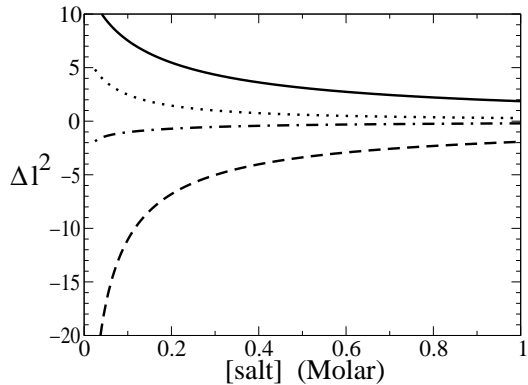}
\end{center}
\end{figure}

\section{Results}\label{sec:results}

We can calculate the free energies of the three interfaces:
solid--salt-solution, solid--protein-crystal and salt-solution--protein-crystal.
This allows us to see whether a protein crystallite, such as a nucleus
or growing crystallite in solution,
is attracted or repelled by the surface. It is attracted if the
free energy change on bringing a protein crystal from the bulk solution
to the surface is negative and repulsive if the change is positive.
When a protein crystallite is brought from the bulk of the solution into
contact with the surface two interfaces are destroyed, 
that between the salt solution and the protein crystal and that between the
salt solution and the solid, and one is created, that between the
solid and the protein crystal. Thus, the free energy change per unit
area of protein crystal brought into contact with the solid is
\begin{equation}
\Delta=-\dsurfsol-\dsurftens+\dsurfxtal,
\label{delta}
\end{equation}
which in the high salt limit is, using Eqs.~(\ref{gams}), (\ref{gamp}) and
(\ref{xtalsol}),
\begin{equation}
\Delta= \frac{\rhosalt\phid^2}{2\kappas}+\sigma\phid~~~\mbox{linear}.
\label{delin}
\end{equation}
The first term comes from the disappearance of the
salt-solution--protein-crystal interface and is positive as the
electrical double layer contribution to the interfacial tension of this
interface is always negative; see Ref.~\cite{sear02} for a discussion.
As $\phid$ has the same sign as that of the charges on the protein
molecules (here taken to be positive), the second term is positive
if the surface and protein molecules have charges of the same sign and
negative if the have opposite signs. As we should expect, surfaces
repel crystals of proteins with the same charge. They attract crystals
of oppositely charged protein molecules, providing that the surface
charge $\sigma$ is sufficiently high. If the charge density is small
or zero then they repel protein crystals due to the free energy cost
of destroying the salt-solution--protein-crystal interface.

\begin{figure}
\begin{center}
\caption{
\lineskip 2pt
\lineskiplimit 2pt
The electrical double layer contribution to the change
in free energy per unit area, times the surface area of the crystal per protein
molecule, when a protein
crystal is brought from the bulk of the salt
solution to the charged surface of a solid, $\Delta l^2$, Eq.~(\ref{delta}).
The solid, dashed and dotted
curves are for salt concentrations of $\rhosalt=0.1$, $0.5$ and
$1\Molar$, respectively.
\label{dsig}
}
\vspace*{0.2in}
\epsfig{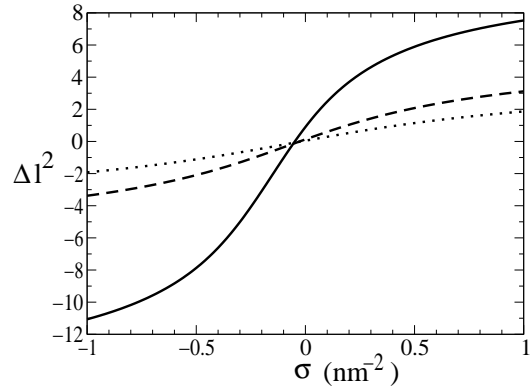}
\end{center}
\end{figure}

Equation (\ref{delin}) is obtained by linearising the \PB~equation.
We have also solved the full equation and present example results
in Figs.~\ref{d} and \ref{dsig}. The parameters for the protein are those used
previously \cite{sear02}, and were chosen to model lysozyme.
This is a small protein with quite a large net charge
for its size, whose behaviour in NaCl solutions
has been extensively studied \cite{muschol95,muschol97,guo99,poon00,warren02}.
The model used here has been to shown to give the
%qualitative nature of the trend and
variation
with NaCl concentration
of the solubility of lysozyme crystals
correctly to within a factor of about 2, it
overpredicts the variation \cite{poon00,warren02}.
The agreement between its predictions and
the surface force apparatus measurements of Sivasankar {\em et al.}
\cite{sivasankar98} of the repulsion between monolayers of
the protein streptavidin, is comparable \cite{sear02}.
To model lysozyme we took a charge $Q=10$ \cite{warren02,muschol97,tanford72},
and a charge density due to the protein $\rhoprot=0.25$nm$^{-3}$,
obtained by taking the volume per lysozyme molecule in the crystal
to be 40nm$^{3}$. The surface area of the protein crystal per lysozyme
molecule is taken to be $l^2=12$nm$^2$ \cite{sear02}.
For proteins such as lysozyme which have a significant net charge
we would expect our predictions to be approximately as accurate as those
of earlier predictions of this theory, i.e., predicting the trends
correctly but giving numbers which are a factor of 2 or more out.
We expect our predictions to be unreliable if the net charge is small,
then the interaction between the dipole moment of the protein and the
surface may dominate, or if the charge is distributed on the surface
in a way which is highly inhomogeneous. If the charge density
within the volume within a few Debye lengths of the surface
differs greatly from $\rho_p$, then our assumption of uniform charge
density due to the protein will be poor. See Ref.~\cite{sear02}
for further discussion of the assumption that the protein crystal
can be modeled by a step-function charge density.
%The surface charge
%density is taken to be either 1 or $0.1$ elementary charges per nm$^2$.

In Fig.~\ref{d} we see that for lysozyme
and solid surfaces with $\sigma=\pm1~\!$nm$^{-2}$, the free energy change
of bringing the protein crystal into contact with the solid is
substantial unless the salt concentration is around $1\Molar$. Even
at a salt concentration of $1\Molar$, the free energy change is
close to $2\kT$ per lysozyme molecule at the surface of the protein
in contact with the solid surface. Whereas at salt concentrations of
about $0.1\Molar$ and lower the free energy change is of order $10\kT$
per lysozyme molecule. Note however, that we have assumed that the
protein crystal is in a salt solution, i.e., a solution with
negligible amounts of protein in it. Decreasing the salt concentration
increases the solubility of lysozyme \cite{muschol97}, and so this
assumption will tend to worsen as the salt concentration drops
\cite{warren02,sear02}. The salt is not only changing the interaction between
the protein and the surface but also the interaction in between protein
molecules and this needs to be borne in mind.

In Fig.~\ref{dsig} we see that at high salt concentrations, $1\Molar$,
the free energy change on bringing a crystal into contact with the
surface is never more than a couple of $\kT$, even for quite highly charged
surfaces. However, for more modest salt concentrations of $0.1\Molar$,
rather large free energy decreases per protein molecule are easily
obtained at modest charges per unit area on the surface. Also, note that
$\Delta$ is slightly positive for an uncharged surface. This is due to
the fact that the electrical double layer contribution to the
protein-crystal--salt-solution interface is negative, Eq.~(\ref{xtalsol}),
and this
contribution is lost when the protein crystal is brought into contact with
the surface.

%Now, of course the difference $\Delta$ is only the electrical double
%layer contribution to the free energy change on bringing a protein crystal
%into contact with the solid surface. There are other contributions, for
%example from dispersion forces. We would expect dispersion forces
%between most solids and protein molecules in water to be attractive
%\cite{israelachvili},
%thus even a neutral surface may attract a protein crystal due
%to these forces.

\begin{figure}
\begin{center}
\caption{
\lineskip 2pt
\lineskiplimit 2pt
The potential, in units of $\kT/e$, as a function of distance $z$ from the
surface. The solid curves are for a salt solution in contact with the
surface and the dashed curves are for a protein crystal against the
surface. The upper solid and dashed curves are for a positively charged
surface, $\sigma=1\,$nm$^{-2}$, and the lower curves are
for a negatively charged surface, $\sigma=-1\,$nm$^{-2}$. The
salt concentration is $0.1\Molar$, and the protein is positively charged,
$\rhoprot=0.25\,$nm$^{-3}$.
\label{phi}
}
\vspace*{0.2in}
\epsfig{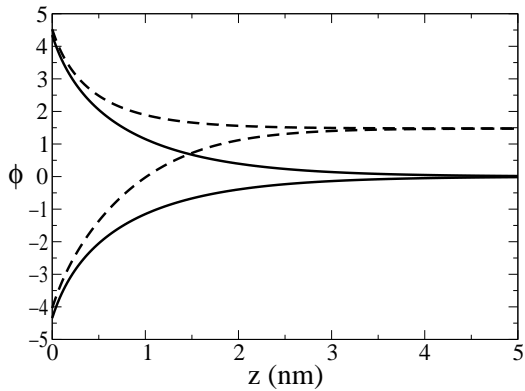}
\end{center}
\end{figure}

The potential near the surface, both when the salt solution is
in contact with the surface and when the protein crystal is, is shown
in Fig.~\ref{phi}. When the protein crystal is in contact with the
surface the potential tends to its value in the bulk of the crystal:
the Donnan potential, which is $1.48$ here. The potential at the surface
is larger than one, and so the linear approximation is quite poor for
the charge $\sigma=\pm1\,$nm$^{-2}$. In particular, Eq.~(\ref{delin})
significantly overestimates the magnitude
of $\Delta$ at high surface charges.
However, Eq.~(\ref{delin}) is reasonably accurate for the smaller charge
per unit area,  $\sigma=\pm0.1\,$nm$^{-2}$. As an
example, for $\sigma=-1\,$nm$^{-2}$ and $\rhosalt=0.1\Molar$,
$\Delta l^2=-11.1$ without linearisation, whereas Eq.~(\ref{delin})
predicts $\Delta l^2=-17.0$.
%For a lower surface charge,
%$\sigma=-0.1\,$nm$^{-2}$ the predictions without and with linearisation is
%$\Delta l^2=-1.02$, i.e., here linearisation is accurate to the first 3
%significant figures.

\section{Conclusion}

We have used an existing \cite{warren02,sear02} generic model of
a protein crystal to estimate the electrostatic contribution $\Delta$ to the
free energy change when a protein
crystal is brought into contact with a surface with a fixed charge
density $\sigma$. Unsurprisingly, the free energy change is negative
for a surface with a charge opposite in sign to that on the protein
molecule, and positive if the surface has a charge with the same sign.
We quantified this and found that very reasonable charge densities,
of 1 elementary charge per nm$^2$ or less, were sufficient to achieve
free energy changes per protein molecule at the surface of several $\kT$
or more, provided the salt concentration was less than around
$1\Molar$. The electrostatic contribution to the free energy
is the energy associated with the electrostatic interactions in between the
surface, protein molecules, and salt and counterions, and the translational
entropy of the salt and counterions.
Typically, there will also be non-electrostatic interactions of the proteins
with the surface, particularly if the surface is at least moderately
hydrophobic. Then the proteins may even unfold resulting in irreversible
adsorption. Thus, our results will be relevant only to those surfaces
which are sufficiently hydrophilic that the proteins interact only
weakly with the surface and do not unfold. However, obtaining such surfaces
is relatively straightforward via the use of self-assembled monolayers
\cite{clark97,ostuni99}.
Our model of the protein is a very simple one, appropriate for
proteins with reasonably large net charges, so that the terms
we calculate should be dominant, and where this charge is
distributed so that the charge density facing the surface is neither
much higher nor much lower than the average charge density
on the surface of the protein.

Perhaps the most important reason for considering protein crystals
at surfaces is interest in their heterogeneous nucleation.
Protein crystals nucleate at
surfaces and so without an understanding
of crystallites at surfaces
we cannot hope to understand how the crystals actually form.
Crystallising proteins is essential for the determination of their
all-important three-dimensional structure via X-ray
crystallography \cite{rosenberger96,chayen02}.
Heterogeneous nucleation is an activated process \cite{debenedetti} and so
proceeds at a rate which scales as $\exp(-\Delta F^*/\kT)$,
where $\Delta F^*$ is the height of the free energy barrier which must
be overcome. Typical sizes of the critical nuclei in nucleation are a few 10s
of protein molecules, and so one face of a crystalline nucleus has perhaps
10, or a few less, protein molecules. The critical nucleus is that
at the top of the barrier, the nucleus which requires a free energy
$\Delta F^*$ to create it \cite{debenedetti}.
Thus, the variation in $\Delta F^*$ with $\Delta$, will be approximately
$10\Delta l^2$. From Figs.~\ref{d} and \ref{dsig}, we see that this
may easily be 10s of $\kT$, causing increases or decreases in the
rate of heterogeneous nucleation
of many orders of magnitude. The
barrier to nucleation may be lowered by 10s of $\kT$, inducing nucleation
in solutions which would otherwise be metastable, or it may be raised
by 10s of $\kT$ preventing nucleation from occurring on one part of the surface.
The former is required if a surface is to be used as a nucleant:
a surface which triggers nucleation \cite{mcpherson88,chayen01}.
The latter is required for spatial control of heterogeneous nucleation
\cite{sanjoh99,sanjoh01}.

Finally, we consider wetting by the crystalline phase of the
surface--salt-solution interface. At coexistence, for a bulk
solution phase in contact with the surface, a direct
surface--salt-solution interface
is not the only possibility. It is possible for a slab of the
protein crystal to be interposed between the surface and the solution
phase. The direct surface--salt-solution interface is then replaced by
by a surface--protein-crystal interface plus solution--protein-crystal interface.
The protein crystal is said to have wet the surface--salt-solution interface.
Wetting
occurs at equilibrium whenever it lowers the free
energy \cite{israelachvili,bonn01}.
Denoting the full interfacial tensions of the surface--salt-solution,
surface--protein-crystal and solution--protein-crystal interfaces
by $\gamma_{s}$, $\gamma_{x}$ and $\gamma$, respectively,
we have that the free energy
change on interposing a slab of protein crystal between the surface and
the solution is $-\gamma_s+\gamma_x+\gamma$. Note that here
the interfacial tension of the solution--protein-crystal interface
appears with a positive sign whereas it appears with a negative sign in
the definition of $\Delta$, Eq.~(\ref{delta}). When a layer
of the protein crystal is created so is a solution--protein-crystal
interface, whereas when an existing protein crystal surrounded
by solution is brought into contact with the surface a
solution--protein-crystal interface is destroyed.

Determining the sign of $-\gamma_s+\gamma_x+\gamma$ is not possible
without determining both the electrostatic and non-electrostatic
contributions to all three
interfacial tensions. However, in Ref.~\cite{sear02}, the interfacial
tension $\gamma$ was estimated to be a few $\kT$ per area of a protein
molecule, i.e., $\gamma l^2=O(1)$. If we further assume that
$-\gamma_s+\gamma_x$ is dominated by electrostatics, then in the
linear regime $-\gamma_s+\gamma_x\simeq\sigma\phid$, from
Eqs.~(\ref{gams}) and (\ref{gamp}). Thus for the electrostatic
attraction between an oppositely charged surface and a protein crystal
to drive wetting of the surface--salt-solution interface, we
require that $|\sigma\phid| l^2 >O(1)$. For highly charged surfaces,
$\sigma$ of order 1nm$^{-2}$, and low salt concentrations, this
inequality is easily satisfied.

If the protein crystal wets the surface--salt-solution interface,
then assuming that the layer of protein crystal can itself nucleate
on the surface it will do so and the bulk protein crystal can grow from
this layer at the surface. Then the barrier to crystallisation will
be close to zero and the protein will readily crystallise
from solution \cite{vlnote}.
This assumes that the layer itself can nucleate, see
Ref.~\cite{bonn01} for an introduction to wetting layers, both
at equilibrium and their nucleation.

%As  $\length^2 \approx 10\,\nm^2$, a surface tension of $1\kT$ $l^2$ area
%is about $0.5\,\milliN\,\metre^{-1}$. For the
%salt-solution--protein-crystal we expect an interfacial tension of perhaps
%$10\kT l^{-2}$.

%Physiological salt concentration is $0.15\Molar$; at this
%salt concentration a protein crystal of a reasonably
%highly charge protein will be strongly attracted to
%(repelled by) a membrane with a high proportion
%of lipids of the opposite (same) sign.

\begin{acknowledgments}

I would like to thank P. Warren for introducing me to the theory
of electrostatic effects in proteins. Work supported by The Wellcome
Trust (069242).

\end{acknowledgments}

%\newpage


\begin{thebibliography}{99}
\bibitem{note} This slightly nonobvious notation is for consistency
with Ref.~\cite{sear02}.
\end{thebibliography}
\end{document}